# Physics-Informed Neural Networks for Modeling the Martian Induced Magnetosphere

Jiawei Gao[1], Chuanfei Dong[1], Chi Zhang[1], Yilan Qin[1], Simin Shekarpaz[1], Xinmin Li[1], Liang Wang[1], Hongyang Zhou[1], Abigail Tadlock[1]

[1]Center for Space Physics and Department of Astronomy, Boston University, Boston, MA, USA

Corresponding author: Jiawei Gao (jgao2@bu.edu) and Chuanfei Dong (dcfy@bu.edu)

**Abstract**

Understanding the magnetic field environment around Mars and its response to upstream solar wind conditions provide key insights into the processes driving atmospheric ion escape. To date, global models of Martian induced magnetosphere have been exclusively physics-based, relying on computationally intensive simulations. For the first time, we develop a data-driven model of the Martian induced magnetospheric magnetic field using Physics-Informed Neural Network (PINN) combined with MAVEN observations and physical laws. Trained under varying solar wind conditions, including $B_{IMF}$, $P_{SW}$, and $\theta_{cone}$, the data-driven model accurately reconstructs the three-dimensional magnetic field configuration and its variability in response to upstream solar wind drivers. Based on the PINN results, we identify key dependencies of magnetic field configuration on solar wind parameters, including the hemispheric asymmetries of the draped field line strength in the Mars-Solar-Electric coordinates. These findings demonstrate the capability of PINNs to reconstruct complex magnetic field structures in the Martian induced magnetosphere, thereby offering a promising tool for advancing studies of solar wind-Mars interactions.

# 1. Introduction

Mars lacks a global intrinsic magnetic field, and thus the interaction between the solar wind and its upper atmosphere generates an induced magnetosphere characterized by complex current systems and dynamic magnetosphere (Acuña et al., 1998; Brain et al., 2017; Dubinin et al., 2011; Nagy et al., 2004; Ramstad et al., 2020; C. Zhang et al., 2025). The induced magnetosphere exhibits a highly variable and asymmetric magnetic environment driven by upstream solar wind conditions. Understanding how the magnetic field is organized and how it responds to upstream solar wind conditions is crucial for studying planetary ion escape and atmospheric evolution driven by the solar wind (Fang et al., 2023; C. Dong et al., 2015a, 2018a; Y. Dong et al., 2019; Luhmann et al., 2004, 2015; C. Zhang et al., 2022).

Over the past decades, both statistical and simulation study have been developed to describe the Martian magnetic environment. Statistical methods, such as slice-averaged magnetic field maps (e.g., Dubinin et al., 2021; He et al., 2021; Rong et al., 2014; C. Zhang et al., 2022), have provided important insights into the general configuration of the induced magnetosphere. However, these approaches face challenges in isolating the effects of specific solar wind drivers. Most statistical studies can only resolve average magnetic field distributions, rather than responses to specific upstream conditions. Simulation methods, such as magnetohydrodynamic (MHD) simulations (e.g., Ma et al., 2004, 2019; C. Dong et al., 2014, 2018b; Sun et al., 2024) and hybrid simulations (e.g., Modolo et al., 2016; Wang et al., 2024), offer a more physically grounded means of modeling Martian induced magnetosphere. However, these simulations are computationally expensive and rely on underlying model assumptions.

For planets with intrinsic magnetic fields such as Mercury, Earth, and Jupiter, empirical magnetospheric models based on basis functions have been widely used to describe the field configuration and its spatial variations (e.g., Korth et al., 2017; Tsyganenko and Sitnov, 2005, 2007; Wilson et al., 2023). These models rely on well-defined basis functions representing current systems (e.g., ring current and tail current), constrained by spacecraft observations (see review by Tsyganenko et al., 2013 and references therein). However, such current basis functions are difficult to apply on Mars. The Martian current systems are irregular, spatially variable, and strongly influenced by upstream conditions, making it difficult to formulate simple basis functions (e.g., Gao et al., 2024; Harada et al., 2025; Ramstad et al., 2020). As a result, magnetospheric modeling methods that work well for magnetized planets cannot be easily applied to unmagnetized planets like Mars.

To overcome these limitations, data-driven machine learning approaches have received growing attention. Recent studies have demonstrated that machine learning can be effectively applied to model Martian crustal magnetic fields (Delcourt et al., 2025) and to separate the induced and crustal magnetic field components (Azari et al., 2023). Among these methods, Physics-Informed Neural Networks (PINNs) are especially promising for modeling magnetic fields in the induced magnetosphere. PINNs incorporate physical laws and boundary conditions directly into the training process (Raissi et al., 2019; Karniadakis et al., 2021; Qin et al., 2023). This allows PINN model to learn from sparse and noisy data while ensuring the solution remains physically consistent (e.g., $\nabla \cdot B = 0$ in this study). Unlike traditional statistical methods, PINNs produce continuous solutions without spatial discretization, making them flexible and efficient for modeling complex magnetic field configurations.

In this study, we apply PINN model to reconstruct the three-dimensional magnetic field in the Martian induced magnetosphere by combining Mars Atmosphere and Volatile EvolutioN (MAVEN) magnetic field observations with physical laws and boundary constraints. By incorporating upstream solar wind parameters, including solar wind dynamic pressure, IMF intensity, and IMF cone angle, the model effectively captures the variability of magnetic field configurations under varying solar wind conditions.

## 2. Data and Methods

### 2.1 Data and Coordinates

We use magnetic field and plasma data obtained by the MAVEN spacecraft from November 1, 2014 to May 15, 2022. The coverage of MAVEN's orbit is shown in Figure 1a–c, illustrating that the spacecraft adequately samples the entire Martian magnetosphere. With an orbital period of approximately 4.5 hours, MAVEN typically observes both the upstream solar wind and the induced magnetosphere within a single orbit (Jakosky et al., 2015). For each orbit, the upstream solar wind conditions are estimated by averaging the upstream parameters measured between two consecutive bow-shock crossings. We select data collected under steady IMF conditions. Out of a total of 13,955 orbits, 3,103 orbits satisfy this criterion (see Appendix A). Magnetic field measurements are provided by the fluxgate magnetometers (MAG) onboard MAVEN, and the vector magnetic field data are resampled at 4s resolution (Connerney et al., 2015). The solar wind velocity and density are obtained from the Solar Wind Ion Analyzer (SWIA) instrument (Halekas et al., 2015).

The upstream solar wind conditions are characterized by solar wind dynamic pressure $P_{SW}$ ($P_{SW} = mNV^2$, $m$ is the proton mass, and $N$ and $V$ are the solar wind density and velocity, respectively), IMF intensity ($B_{IMF}$), and IMF cone angle $\theta_{cone}$ (the angle between the IMF and the Mars–Sun direction), which serve as external control parameters in the data-driven model. These parameters are chosen because they are known to exert the strongest influence on the topology of the Martian induced magnetosphere (Zhang et al., 2022; Fang et al., 2023). The histograms of these upstream parameters are shown in Figure 1d–f to illustrate their statistical distribution.

Since the magnetic field configuration of the induced magnetosphere is controlled by the IMF orientation (e.g., Zhang et al., 2022; Zhang et al., 2025), we analyze the magnetic field distribution in the Mars–Solar–Electric (MSE) coordinates. In MSE coordinates, the X-axis is anti-parallel to the solar wind flow, the Z-axis is aligned with the solar wind electric field ($\mathbf{E}_{SW} = -\mathbf{v}_{SW} \times \mathbf{B}_{IMF}$, where $\mathbf{v}_{SW}$ is the solar wind flow). The Y-axis completes the right-handed coordinate system. In this study, the effects of crustal magnetic fields on the induced magnetosphere are not considered, and data strongly influenced by crustal fields are excluded. Specifically, we retain only data points for which the crustal field magnitude satisfies $|B_{crustal}|$<10 nT, where $B_{crustal}$ is the magnetic field strength calculated by a crustal fields model (Gao et al., 2021). After applying these selection criteria, the final dataset contains 1.61 million data points. For each point, we record the upstream solar wind parameters ($P_{SW}$, $B_{IMF}$, $\theta_{cone}$), spacecraft position, and magnetic field measurements in the MSE coordinates.

## 2.2 Physics-Informed Neural Network Framework

We adopt Physics-Informed Neural Networks (PINNs) to reconstruct the steady-state three-dimensional magnetic field distribution around Mars. The network takes spatial coordinates $(x, y, z)$ and selected upstream solar wind parameters as input, and outputs the three magnetic field components $(B_x, B_y, B_z)$. The PINN architecture is implemented as a fully connected feed-forward neural network with three hidden layers of 128 neurons each (Figure 1g). The hyperbolic tangent activation function (tanh) is used in all hidden layers, and all derivatives required by the physics constraints are computed using automatic differentiation (Paszke et al., 2017), following standard PINN methodology (Raissi et al., 2019; Karniadakis et al., 2021, Wang et al., 2023).

The PINN model is trained by minimizing composite loss function that incorporates observational data loss, physical loss, and boundary loss. The total loss is defined as

$$L_{total} = L_{data} + \lambda_{phys}L_{phys} + \lambda_{bc}L_{bc} \tag{1}$$

$\lambda_{phys}$ and $\lambda_{bc}$ denotes the loss weight. The data loss enforces agreement between the predicted magnetic field and the MAVEN observed magnetic field:

$$L_{data} = \frac{1}{N_d}\sum_{i=1}^{N_d}\left|\boldsymbol{B}_i^{obs} - \boldsymbol{B}_i^{model}\right|^2 \tag{2}$$

where $B_i^{obs}$ is the observed magnetic field, $B_i^{model}$ is the predicted magnetic field, and $N_d$ is the number of data points. To ensure that the model remains physically consistent, we enforce the divergence-free constraint of the magnetic field ($\nabla \cdot B = 0$), thus:

$$L_{phys} = \frac{1}{N_d}\sum_{i=1}^{N_d}\left|\frac{\partial B_x}{\partial x} + \frac{\partial B_y}{\partial y} + \frac{\partial B_z}{\partial z}\right|_i^2 \tag{3}$$

Because the induced magnetosphere is shaped by both the solar wind and the planetary obstacle, boundary conditions are incorporated into both the upstream boundary loss and the surface boundary loss ($L_{bc} = L_{bc,up} + L_{bc,surf}$).

At upstream of Mars (e.g., $x = 3R_m$, $R_M$=3393.5 km), the magnetic field approaches the IMF. In the MSE coordinates, the IMF consistently exhibits a positive Y component, and its magnitude is the upstream parameters $B_{IMF}$. We impose $\boldsymbol{B}(x = 3R_m) = \boldsymbol{B}_{IMF}$, leading to the boundary loss:

$$L_{bc,up} = \frac{1}{N_u}\sum_{i=1}^{N_u}\left|\boldsymbol{B}(x_i) - \boldsymbol{B}_i^{IMF}\right|^2 \tag{4}$$

Previous observations confirm that the external magnetic field at ionospheric altitudes is predominantly horizontal (Mittelholz et al., 2017; Gao et al., 2024; Harada et al., 2025). Therefore, the normal component of the magnetic field vanished at the Mars surface ($r = R_M$), preventing magnetic field lines from crossing the planetary body. The corresponding loss is

$$L_{bc,surf} = \frac{1}{N_s}\sum_{i=1}^{N_s}|\boldsymbol{B}(x_i) \cdot \boldsymbol{n}_i|^2 \tag{5}$$

where $\boldsymbol{n}_i$ is the outward-pointing unit normal vector at that point. At the Martian surface, we uniformly sample $N_s = 4096$ points on the sphere at every training iteration. All input quantities are normalized using Z-score standardization, so that each input dimension has similar scale and numerical range (Wang et al., 2023). This improves network stability and accelerates convergence. Because the spatial coordinates are standardized before computing derivatives, the divergence-free constraint includes a correction to account for the rescaled spatial gradients, via the chain rule (Delcourt et al., 2025).

In this study, we train a total of four models. Models PINN-A1, PINN-A2, and PINN-A3 use the same input parameters ($P_{SW}$, $B_{IMF}$) but differ in their loss-function. Model PINN-A1 incorporates all three loss terms in the loss function ($L_{total} = L_{data} + \lambda_{phys}L_{phys} + \lambda_{bc}L_{bc}$), while PINN-A2 includes only the data and physics loss terms ($L_{total} = L_{data} + \lambda_{phys}L_{phys}$). PINN-A3 relies solely on the data loss term ($L_{total} = L_{data}$). These three models serve as an ablation study to examine the influence of different loss terms. Models PINN-B use a single upstream parameter $\theta_{cone}$ as input. Since the amount of observational data is limited, incorporating too many input parameters in one model may increase model complexity, hinder convergence, and reduce accuracy and generalization performance (Bonfanti et al., 2024). We also attempted to train models using a larger set of input parameters, such as $P_{SW}$, $B_{IMF}$, and $\theta_{cone}$, but found that the models struggled to converge and failed to produce meaningful results. Therefore, the PINN-B model is proposed to evaluate the individual contribution of $\theta_{cone}$ to the Martian induced magnetosphere.

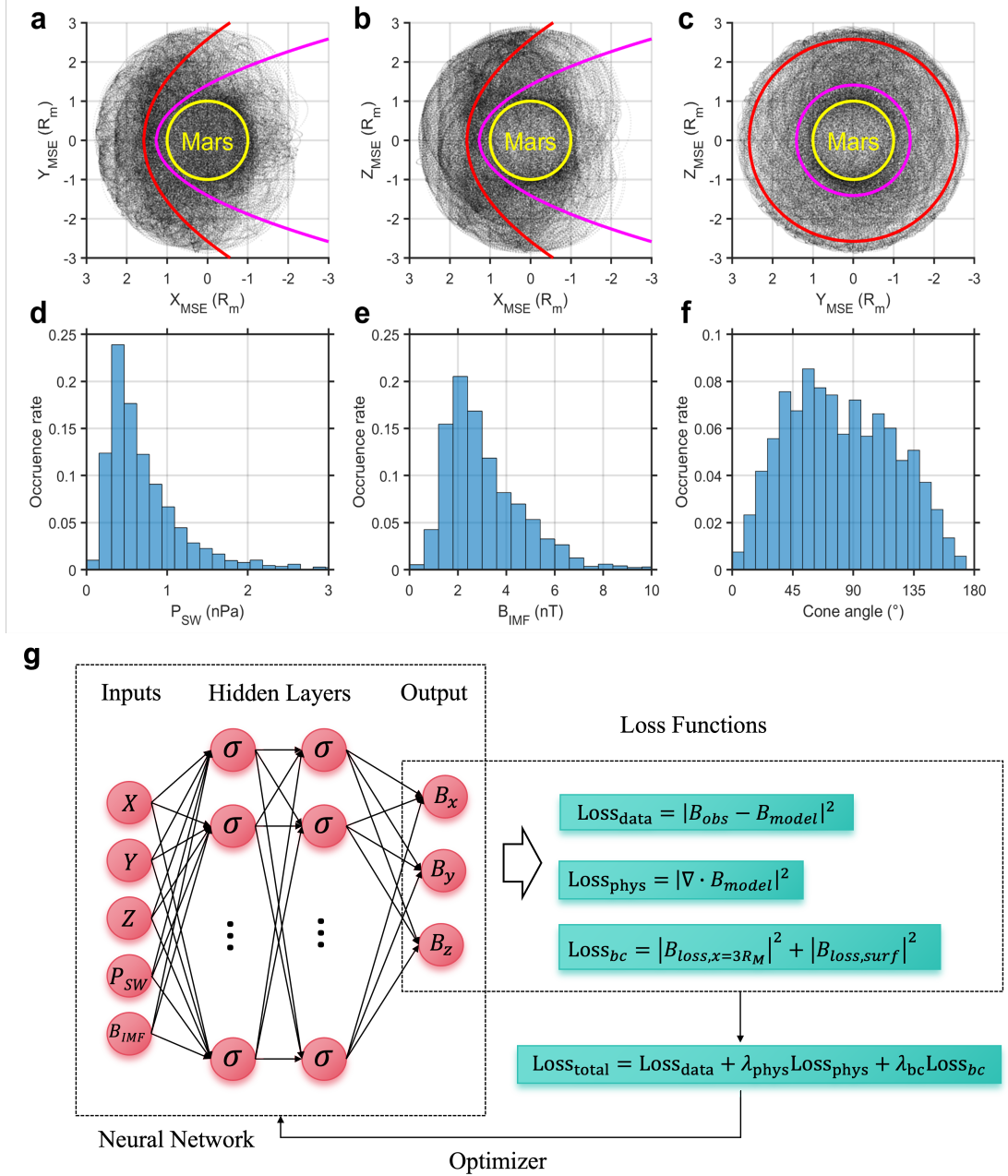

**Figure 1. Training dataset and Physics-Informed Neural Network (PINN) architectures.** (a-c) MAVEN orbit coverage of the training dataset projected onto the (a) $XY_{MSE}$, (b) $XZ_{MSE}$, and (c) $YZ_{MSE}$. (d-f) Histograms of (d) solar wind dynamic pressure $P_{SW}$, (e) the interplanetary magnetic field (IMF) strength, and (f) the IMF cone angle. (g) The framework of PINN-A1 architecture. $\sigma$ denotes the trainable neural network parameters optimized through the loss function. For model PINN-B, the input parameters $P_{SW}$, $B_{IMF}$ are replaced with the IMF cone angle.

## 2.3 Training history and Model Performance

All PINN models were trained for 500 epochs to allow direct comparison, with both weight $\lambda_{phys}$ and $\lambda_{bc}$ set to 1 (see Appendix B). Training was conducted using batches of $5\times10^5$ samples.

The full dataset was randomly split, with 80% used for training and 20% for validation. We do not include a separate test dataset, unlike some machine learning studies, because the observational data are limited in their spatial and upstream condition coverage. A test dataset with insufficient coverage would not yield meaningful or generalizable results. This approach is consistent with many PINN-based studies where, due to limited data availability, a traditional test set is not always used (Raissi et al., 2019).

The training and validation loss curves are shown in Figure 2. Across all models, both training and validation losses decrease rapidly during the first ~150 epochs and continue to converge gradually afterward. Notably, Figure 2c shows that while model PINN-A3 achieves a slightly lower training loss than PINN-A1, it suffers from a significantly larger physics loss ($\nabla \cdot B$), indicating clear overfitting in the absence of physical constraints.

We further evaluate model performance using the root-mean-square (RMS) misfit. With the inclusion of physical and boundary loss terms, PINN-A1 exhibits substantially reduced physics and boundary losses compared to PINN-A2 and PINN-A3 (see Table 1 in Appendix B), while maintaining comparable data fitting accuracy. The spatial distribution of model misfit across the entire dataset is shown in Figure 2e–g. Misfits are generally higher near the planet and lower outside the bow shock. This pattern is likely due to two factors. First, although the upstream IMF is assumed to be steady within a single orbit, its intensity and direction vary continuously, leading to deviations between the induced magnetic field and the real-time upstream IMF orientation. These deviations are relatively small in the magnetosheath but become more pronounced closer to Mars (Y. Dong et al., 2019; Cheng et al., 2025). Second, although we only use data from non-crustal field regions, the influence of Martian crustal fields interacting with the solar wind, especially on the dayside, cannot be fully eliminated (Renzaglia et al., 2023; Harada et al., 2025).

The correlation maps between observations and model predictions, along with the histograms of data residuals (see Figure 6 and Figure 7 in Appendix C), show that most data points are centered around the diagonal across all regions, indicating that most observations are well reproduced by the model. Additionally, the residuals for all three magnetic field components exhibit a symmetric distribution centered around zero, suggesting the absence of systematic bias.

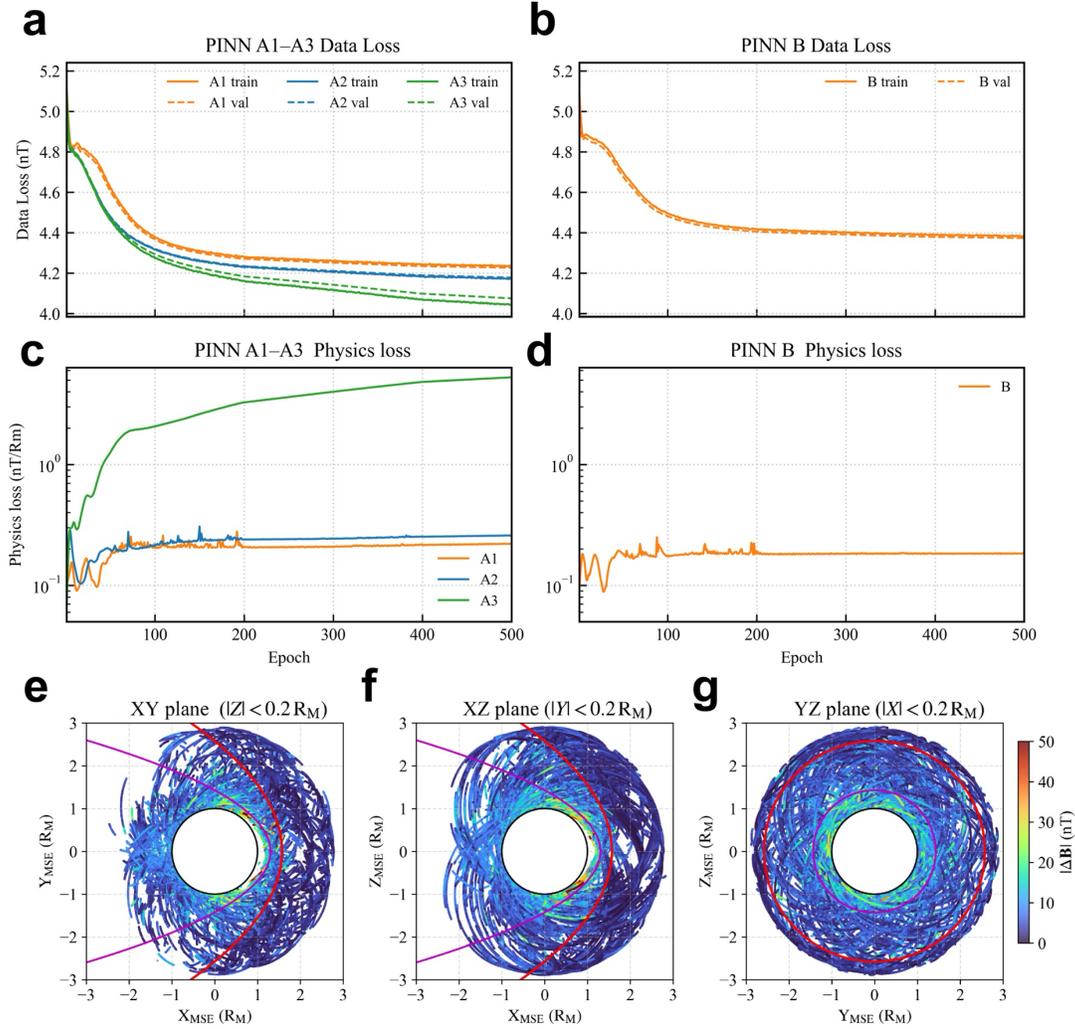

**Figure 2. Training history of the PINN models and the misfit distribution.** (a) Total training loss and validation loss for PINN A1-A3. Solid lines denote training loss and dashed lines denote validation loss. (b) Same as (a), but for PINN-B model. All curves represent root-mean-square (RMS) misfit in units of nanotesla (nT) B intensity relatively detB. (e-f) Model misfit distribution from PINN-A1, shown in the (e) $XY_{MSE}$, (f) $XZ_{MSE}$, and (g) $YZ_{MSE}$ plane, respectively.

## 3 Results

### 3.1 PINN-Reconstructed Global Magnetic Field Structure

Figure 3 presents the magnetic field distribution from the PINN-A1 model on $XY_{MSE}$, $XZ_{MSE}$, and $YZ_{MSE}$ slices. Compared with the corresponding results obtained from MAVEN observations (Figure 8 in Appendix C), the model shows good agreement, indicating a successful reconstruction of the induced magnetosphere. The model focuses on reconstructing the magnetic field structure within the Martian bow shock, and the weak and turbulent magnetic fields in the foreshock region are not addressed in this study (Jarvinen et al., 2022; Tadlock et al., 2025). The magnetic field outside the bow shock are replaced with the corresponding upstream IMF as inputs (Figure 9 in Appendix C). Compared with the PINN-A2 and PINN-A3 models (Figure 10 and Figure 11 in Appendix C), the PINN-A1 model avoids the small-scale irregularities seen in PINN-A3 and prevents the unphysical magnetic field lines crossing the Martian body observed in PINN-A2, showing the best overall performance. Therefore, we focus on the PINN-A1 model in the following analysis.

The PINN-A1 model successfully reproduces the three-dimensional configuration of the Martian induced magnetosphere across a wide range of upstream solar wind conditions. The model captures key features such as the draping of the IMF around the planet (Figure 3b), the magnetic barrier on the dayside, and the extension of the draped IMF into the magnetotail, forming tail lobes (e.g., McComas et al., 1986; Rong et al., 2014; Zhang et al., 2022). An asymmetry in magnetic field draping is observed between the $+Z_{MSE}$ and $-Z_{MSE}$ hemispheres: IMF draping is stronger in the $+Z_{MSE}$ hemisphere on the dayside (Figure 3e) and $B_y$ becomes even negative near the polar region of the $-Z_{MSE}$ hemisphere. The $-B_y$ component in the magnetotail, a well-known feature reported in previous studies (Du et al., 2013; T. Zhang et al., 2010; Dubinin et al., 2021), is notably enhanced in two regions. One appears near the planet at the south pole and another in the far tail. In the $YZ_{MSE}$ plane, a looping like magnetic field structure around Mars is observed, appearing already on the dayside (Figure 12 in Appendix C) (Chai et al., 2016; 2019), characterized by a clockwise rotation of the magnetic field when viewed from the Sun toward Mars.

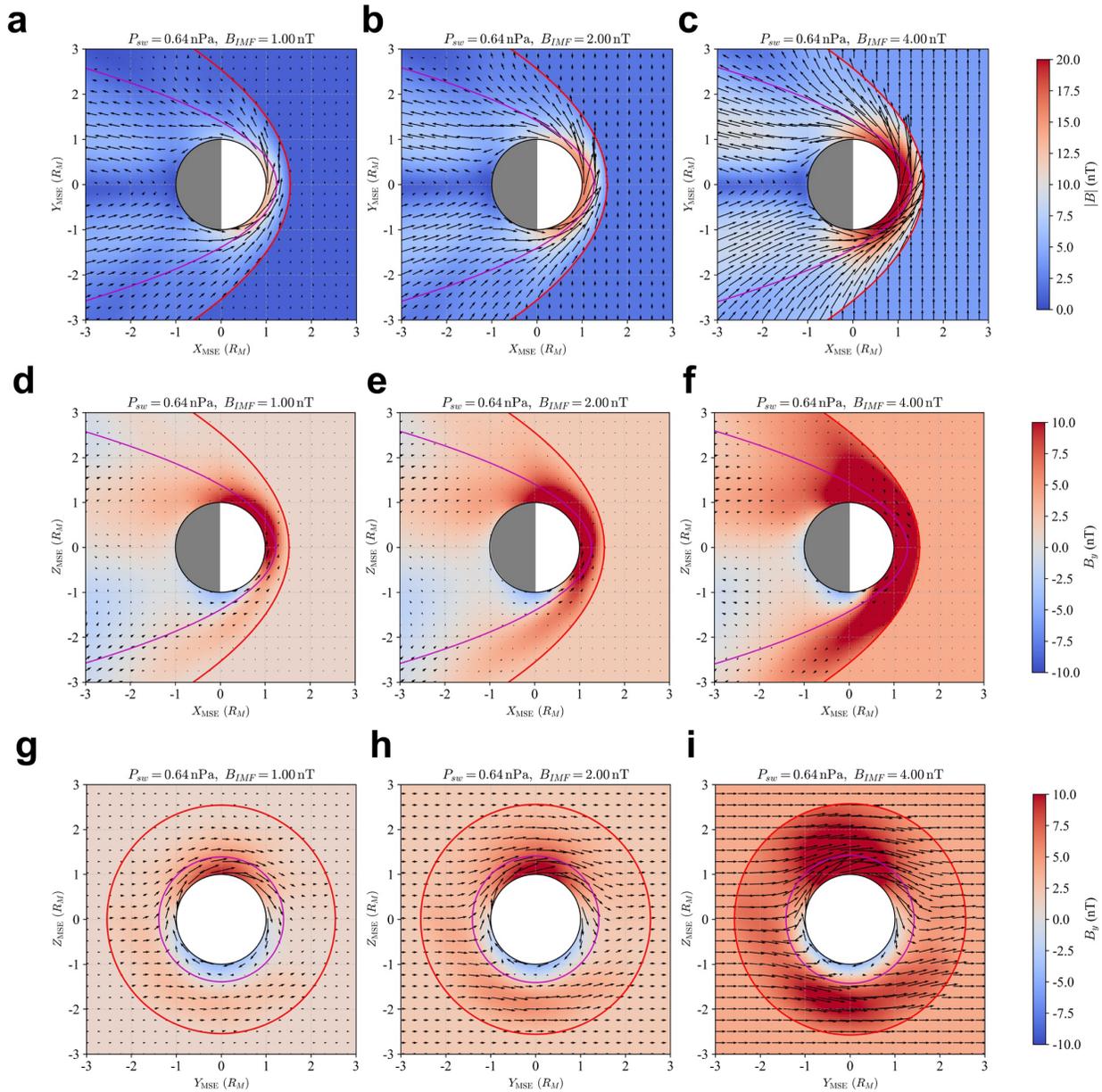

**Figure 3. Magnetic field distribution in the induced magnetosphere under varying upstream IMF strengths from the PINN-A1 model.** Magnetic field vectors are shown in the slices of the (a-c) $XY_{MSE}$, (d-f) $XZ_{MSE}$, and (g-i) $YZ_{MSE}$ planes, respectively. Panels from left to right correspond to upstream $B_{IMF}$ of 1 nT, 2 nT, and 4 nT. The $P_{SW}$ is fixed at 0.64 nPa in all cases. The red and magenta lines denote the shape of the bow shock and the magnetic pileup boundary (MPB) (Němec et al., 2020). Note that in the $XY_{MSE}$ plane, the color bar represents the magnetic field intensity, whereas in the $XZ_{MSE}$ and $YZ_{MSE}$ planes, it represents the $B_y$ component.

## 3.2 Dependence on $B_{IMF}$ and $P_{SW}$

Both $B_{IMF}$ and $P_{SW}$ are widely believed to significantly influence the magnetic structure of an induced magnetosphere (e.g., T. Zhang et al., 1994; C. Zhang et al., 2022). To further examine their roles, we compare the magnetic field configuration under different upstream conditions. **Figure 3** shows the dependence on $B_{IMF}$ by fixing $P_{SW}$ at 0.64 nPa while varying $B_{IMF}$ at 1 nT, 2 nT, and 4 nT. **Figure 4** shows the effect of $P_{SW}$ by fixing $B_{IMF}$ at 2 nT and varying $P_{SW}$ at 0.2 nPa, 0.64 nPa, and 1.6 nPa. The chosen fixed parameters ($P_{SW}$ = 0.64 nPa, $B_{IMF}$= 2 nT) represent the median values of upstream IMF conditions (Liu et al., 2021).

We find that the intensity of the induced magnetic field is most sensitive to variations in $B_{IMF}$. As $B_{IMF}$ increases, the magnetic field intensity in the magnetosheath, magnetic barrier, and magnetotail increases significantly (**Figure 3**a–c). The IMF draping becomes more pronounced, with its spatial extent along the $Z_{MSE}$ direction expanding from about 1 $R_M$ at $B_{IMF}$=1 nT to about 2 $R_M$ at $B_{IMF}$=4 nT.

When $B_{IMF}$ is fixed and $P_{SW}$ increases (**Figure 4**), the magnetic field strength in the magnetotail increases only slightly, but the dayside magnetic barrier becomes more compressed and shifts closer to the planetary surface (**Figure 4**c). In the $YZ_{MSE}$ plane, it is clear that magnetic field intensities near the bow shock become weak under high $P_{SW}$ (**Figure 4**i), consistent with a more compressed magnetosheath. Interestingly, the -$B_y$ component around the Martian south pole remains consistent across all upstream conditions, regardless of changes in either $B_{IMF}$ and $P_{SW}$.

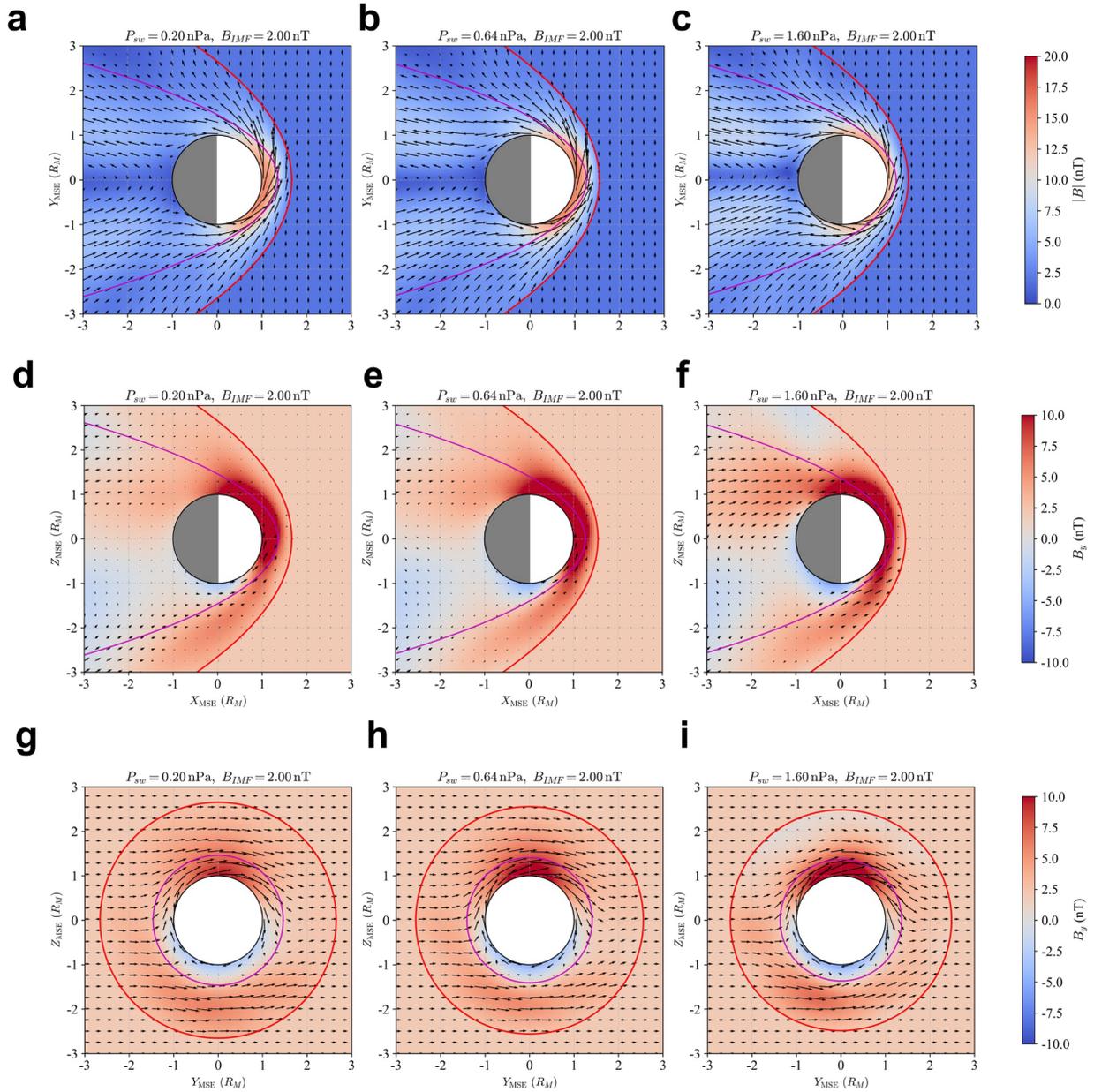

**Figure 4. Magnetic field distribution in the induced magnetosphere under varying upstream solar wind dynamic pressure from the PINN-A1 model.** Magnetic field vectors are shown in the slices of the (a-c) $XY_{MSE}$, (d-f) $XZ_{MSE}$, and (g-i) $YZ_{MSE}$ planes, respectively. Panels from left to right correspond to upstream $P_{SW}$ of 0.2 nPa, 0.64 nPa, and 1.6 nPa. The $B_{IMF}$ is fixed at 2 nT in all cases.

### 3.3 Dependence on IMF cone angle

The magnetic field draping configuration in the induced magnetosphere is strongly modulated by the IMF cone angle (e.g., Delva et al., 2017; Rong et al., 2016; C. Zhang et al., 2022). A kink-like magnetic field structure, known as the inverse polarity reversal layer (IPRL), has been observed in quasi-parallel magnetosheath, where the IMF is nearly aligned with the bow shock normal (Romanelli et al., 2015; C. Zhang et al., 2022; Tadlock et al., 2025). Previous studies suggest that the IPRL disappears when the IMF cone angle approaches 90°, but emerges when the cone angle is close to 0° or 180°.

Figure 5 presents the magnetic field configuration under IMF cone angles of 30°, 90°, and 150° from the PINN-B model. Our results confirm the appearance of the IPRL when the cone angle deviates significantly from 90°, in quasi-parallel magnetosheath. In such conditions, the magnetic field intensity shows a local minimum near the magnetic pileup boundary (MPB), and the model reproduces a smooth field transition across the bow shock (**Figure 5**a, c). Interestingly, we find that the strength and symmetry of magnetic field draping vary with cone angle. At IMF cone angle of 30° (150°), the draping field in the magnetic barrier is stronger in the $-Y_{MSE}$ ($+Y_{MSE}$) hemisphere, respectively. When the cone angle is 90°, the draping pattern becomes symmetric between the $\pm Y_{MSE}$ hemispheres and the overall magnetic field strength is maximized.

While previous studies have proposed that the induced magnetosphere may degenerate under very small IMF cone angles (Q. Zhang et al., 2024), our PINN-B model results indicate that a structured induced magnetosphere still forms even at extreme cone angles. The model consistently shows magnetic field draping on the dayside, a well-developed magnetotail, as well as looping magnetic field, across all cone angle cases, although the magnetic field strength in the magnetic barrier is reduced when the cone angle approaches 0° or 180°.

The role of the IMF $B_x$ component in shifting the magnetotail current sheet remains debated. Some studies suggest that $B_x$ can cause significant displacement of the current sheet (e.g., McComas et al., 1986; Wen et al., 2025), while others report minimal dependence on $B_x$ (Romanelli et al., 2015; Rong et al., 2016). In our model, the magnetotail current sheet remains nearly centered across all cone angles, including at 0° and 180°. The displacement induced by the IMF $B_x$ component is found to be less than 0.2 $R_M$ along the $Y_{MSE}$, suggesting that its influence is minor.

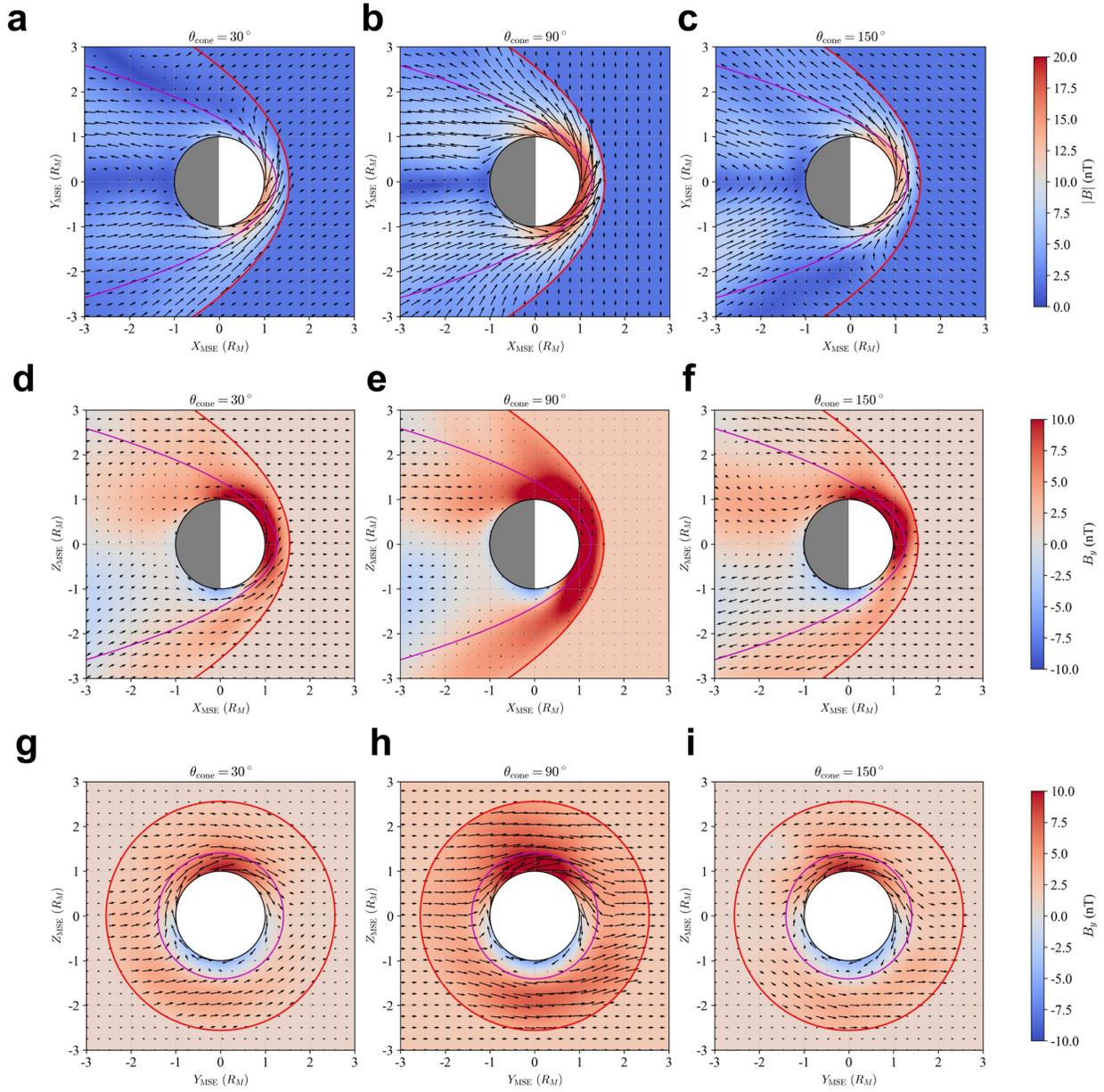

**Figure 5. Magnetic field distribution in the induced magnetosphere under varying IMF cone angle from the PINN-B model.** Magnetic field vectors are shown in the slices of the (a-c) $XY_{MSE}$, (d-f) $XZ_{MSE}$, and (g-i) $YZ_{MSE}$ planes, respectively. Panels from left to right correspond to IMF cone angle 30°, 90°, and 150°.

**Discussion and Conclusion**

    Using magnetic field data collected by MAVEN, we developed three-dimensional magnetic field models of the Martian induced magnetosphere based on Physics-Informed Neural Networks

(PINNs). We also demonstrated that PINN models (PINN-A1, PINN-B) can be effectively applied to investigate the response of the induced magnetosphere to upstream solar wind conditions, including $P_{SW}$, $B_{IMF}$, and $\theta_{cone}$. The main findings are summarized as follows:

1. The intensity of the draped magnetic field in the induced magnetosphere is primarily controlled by $B_{IMF}$, while $P_{SW}$ governs the compression of the magnetic barrier. IMF draping is stronger in the $+Z_{MSE}$ hemisphere on the dayside and becomes negative near the southern pole in the $-Z_{MSE}$ hemisphere.

2. The $-B_y$ component in the magnetosphere is enhanced in two regions: near the south pole close to the planet and in the distant magnetotail. The $-B_y$ component around the Martian south pole remains consistent across all upstream conditions, regardless of variations in $B_{IMF}$, $P_{SW}$, and $\theta_{cone}$.

3. The strength and symmetry of magnetic field draping vary with the cone angle. When the IMF cone angle is 30° (150°), the draping field lines in the magnetic barrier are stronger in the $-Y_{MSE}$ ($+Y_{MSE}$) hemisphere, respectively.

4. The influence of the IMF $B_y$ component on shifting the magnetotail current sheet is minimal. The current sheet remains nearly centered across all cone angles, including at 0° and 180°, with any displacement limited to less than 0.2 $R_M$ along the $Y_{MSE}$ axis.

Together, these results demonstrate that $B_{IMF}$ and $P_{SW}$ regulate the induced magnetosphere through complementary but distinct processes. The upstream IMF strength determines the magnitude of the draped magnetic field around the planet on the dayside (Azari et al., 2023). Meanwhile, an increase in $P_{SW}$ reduces the subsolar standoff distances of both the bow shock (Němec et al., 2020), MPB (Matsunaga et al., 2017) and the ionopause (Chu et al., 2021), thereby leading to a stronger compression of the magnetic barrier.

The draped magnetic field in the Martian magnetosphere is expected to align in the $+Y_{MSE}$ direction. However, the presence of a significant $-B_y$ component has attracted considerable attention (e.g., Chai et al., 2016; Du et al., 2013; Dubinin et al., 2021; Rong et al., 2014; C. Zhang et al., 2022; T. Zhang et al., 2010). Several mechanisms have been proposed to explain the origin of this $-B_y$ component, including current systems (Chai et al., 2016; Ramstad et al., 2020), wrapped more tightly (T. Zhang et al., 2010), and the bending of draped field lines associated with plume dynamics (Chai et al., 2019; Dubinin et al., 2019). Our models also reveal a $+B_z$ ($-B_z$) component in the $-Y_{MSE}$ ($+Y_{MSE}$) hemisphere and enhanced $+B_y$ in the $+Z_{MSE}$ hemisphere around the planet (Figure 3h). These features are more consistent with a looping magnetic field encircling the planet.

Based on our model results, we suggest that the observed looping magnetic structure arises from a combined effect of plume dynamics and current systems. The ionospheric plasma is picked up and accelerated by solar wind in the +E direction, producing a recoil effect in the –E direction. This process forms a dense, low-velocity planetary ion trail (Dubinin et al., 2019; Inui et al., 2019; C. Zhang et al., 2022, 2025). The plume, which exhibits higher ion density in the +E hemisphere, leads to stronger magnetic field draping, as seen in the $XZ_{MSE}$ plane (Figure 3e). The ion trail provides a direct connection between the Martian ionosphere and the solar wind, facilitating the excavation of ionospheric electrons (Ramstad et al., 2020). Given the much larger gyroradii of planetary ions, tailward electrons establish a sunward current confined to the MPB. In the ionosphere, electrons gyrating around draped magnetic field lines and the tailward moving of planetary ions contribute to a tailward current, which has been observed at ~150 km altitude in the terminator region (Gao et al., 2024). These sunward and tailward currents form a closed current system that produces a clockwise looping magnetic field when viewed from the Sun toward Mars, consistent with the mechanism proposed by Chai et al. (2016). If this sunward/tailward current system is indeed responsible for the observed $-B_y$, then the current density must increase under higher $B_{IMF}$ conditions to maintain a relatively constant $-B_y$ magnitude across all upstream conditions. This hypothesis can be examined in the future through observations and simulations.

The PINN model successfully reconstructs the Martian induced magnetosphere; however, several limitations remain. First, since the crustal magnetic fields are not included in the model, their influence on the magnetosphere, such as twisting of the magnetotail (DiBraccio et al., 2018, 2022), cannot be captured in this model. Second, the model is constructed in the MSE coordinates, which neglects the influence of the IMF clock angle. Building future models in the MSO frame would enable investigation of IMF clock angle effects. Third, upstream parameters not included in this study, such as solar extreme ultraviolet (EUV) flux and Martian seasonal variations ($L_s$), can also affect the structure of the induced magnetosphere (e.g., C. Dong et al., 2015b; Fang et al., 2023; Gao et al., 2024). Future work should incorporate additional upstream drivers using more advanced modeling techniques (e.g., Raissi et al., 2024; Liu et al., 2025) to better understand their individual effects. Moreover, the recent launch of the EscaPADE mission (Lillis et al., 2024), together with the ongoing Tianwen-1 (Wan et al., 2020) and MAVEN missions, will enhance real-time monitoring of upstream solar wind conditions at Mars. This observational capability will enable more accurate modeling of the temporal variability of the Martian space environment, which is a key focus for future research.

In conclusion, our results demonstrate that a PINN can successfully reconstruct the three-dimensional magnetic field structure of the Martian induced magnetosphere. This approach provides a powerful tool for integrating sparse observational data with physical laws, paving the way for a more comprehensive understanding of solar wind-Mars interactions.

**Appendix A**

**Coordinate transformation for MAVEN data**

The PINN model is constructed using data organized in the Mars–Solar–Electric (MSE) coordinates, while the original MAVEN measurements are provided in the Mars–Solar–Orbital (MSO) coordinates, where the X-axis points from Mars toward the Sun, the Z-axis points toward orbital north, and the Y-axis is opposite to Mars orbital motion. In MSE coordinates, X-axis is anti-parallel to the solar wind flow, the Z-axis aligns with the solar wind electric field, $\boldsymbol{E}_{SW} = -\boldsymbol{v}_{SW} \times \boldsymbol{B}_{IMF}$, where $\boldsymbol{v}_{SW}$ is the solar wind flow and $\boldsymbol{B}_{IMF}$ is the interplanetary magnetic field (IMF).

Following the procedure used in earlier studies (e.g., Rong et al., 2014; Liu et al., 2021; Zhang et al., 2022), we first determined the upstream solar wind conditions for each orbit before performing the MSO to MSE rotation. From 13,955 MAVEN orbits, 8,742 bow-shock crossings satisfied the requirement that the time interval between the inbound and outbound bow-shock crossings exceeded 1.5 hours, and the bow-shock crossing times were identified by manual inspection of MAVEN observations. The directions of the solar wind flow and the IMF were obtained from 25-minute averages of solar wind data starting 5 minutes after the outbound and 5 minutes before the inbound bow-shock crossings. The upstream values for each orbit were then defined as the average of the two measurements, with the solar wind velocity given by $\mathbf{v}_{SW} = \frac{\mathbf{v}_{SW1} + \mathbf{v}_{SW2}}{2}$, solar wind density given by $\mathbf{N}_{SW} = \frac{\mathbf{N}_{SW1} + \mathbf{N}_{SW2}}{2}$, and the IMF by $\mathbf{B}_{IMF} = \frac{\mathbf{B}_{IMF1} + \mathbf{B}_{IMF2}}{2}$. To ensure a stable IMF orientation for defining the MSE coordiante, we retained only those orbits for which the angle between $\boldsymbol{B}_{IMF1}$ and $\boldsymbol{B}_{IMF2}$ less than 30 degrees. Because upstream IMF conditions are constantly changing, the induced magnetosphere response to short-period IMF variations (<4.5-hour, period of the MAVEN orbit) are smoothed out in the statistics (Ramstad et al., 2020).

**Appendix B**

**Training details and hyper-parameter selection for the PINN model**

A total of four neural network models were trained using the PyTorch library (Paszke et al., 2017) on an NVIDIA GeForce GTX 5070 Ti GPU. Models PINN-A1, PINN-A2, and PINN-A3 share the same input parameters ($P_{SW}$, $B_{IMF}$) but differ in their loss-function configurations. Models PINN-B use a single upstream parameter $\theta_{cone}$ as input. We also attempted to train models using a larger set of input parameters, such as $P_{SW}$, $B_{IMF}$, and $\theta_{cone}$, but found that the models struggled to converge and failed to produce meaningful results.

All models share the same PINN architecture and hyperparameters; the only difference lies in the loss functions for the PINN A1–A3 models. Training is performed using the Adam optimizer with an initial learning rate of 0.001. A learning rate scheduler (StepLR) reduces the learning rate by a factor of 0.5 every 200 epochs, and early stopping based on the validation loss is applied to prevent overfitting (Wang et al., 2023). Mini-batches of $5 \times 10^5$ samples are used during training to ensure both efficient GPU utilization and sufficient statistical results within each batch. Network parameters are initialized using Xavier initialization to enhance training stability (Wang et al., 2023).

To balance the components of the composite loss function, we assign fixed loss weights: $\lambda_{phys} = 1$ for the physical loss and $\lambda_{bc} = 1$ for the boundary loss. We also tested smaller weights (e.g., 0.1 and 0.01), but found that the physics loss continued to grow during training. Conversely, larger weights (e.g., 10 and 100) overly smoothed the magnetic field configuration and caused the model to fail in accurately reproducing the induced magnetosphere. Although a variety of adaptive weighting schemes have been proposed (e.g., Raissi et al., 2024, and references therein), we adopt fixed manual weights in this study to allow for direct comparison across different models. The model configurations and root-mean-square (RMS) misfit for all PINN models at the final training epoch is shown in Table 1.

**Table 1. Model configurations and root-mean-square (RMS) misfit for all PINN models.** The table lists the training loss, validation loss, physics loss (∇·B), and boundary loss evaluated at the final training epoch.

| Model | Input parameters | Loss terms | Data loss | | Physics loss | Upstream boundary loss (nT) | Surface boundary loss (nT) |
|---|---|---|---|---|---|---|---|
| | | | Train loss (nT) | Validation loss (nT) | | | |
| PINN-A1 | x, y, z, $P_{SW}$, $B_{IMF}$ | $L_{data} + L_{phy} + L_{bc}$ | 4.23 | 4.23 | 0.22 | 0.12 | 0.26 |
| PINN-A2 | x, y, z, $P_{SW}$, $B_{IMF}$ | $L_{data} + L_{phy}$ | 4.17 | 4.17 | 0.25 | 1.04 | 1.98 |
| PINN-A3 | x, y, z, $P_{SW}$, $B_{IMF}$ | $L_{data}$ | 4.04 | 4.07 | 5.26 | 1.24 | 2.46 |
| PINN-B | x, y, z, $\theta_{cone}$ | $L_{data} + L_{phy} + L_{bc}$ | 4.31 | 4.31 | 0.20 | 0.12 | 0.18 |

a. The physics loss has units of nT/$R_m$ and therefore is not directly comparable to the other loss terms.

## Appendix C

### Performance evaluation of PINN Models

Figure 6 presents the correlation between MAVEN magnetic field observations and the predictions from the PINN-A1 model. Histograms of data residuals across different regions are shown in Figure 7. Magnetic field distribution in the Martian induced magnetosphere under varying upstream IMF strengths from MAVEN observation is shown in Figure 8. Figure 9 illustrates the magnetic field distribution in the Martian induced magnetosphere under varying upstream IMF strengths, based on the PINN-A1 model without IMF replacement. Similar results from the PINN-A2 and PINN-A3 models are shown in Figure 10 and Figure 11, respectively. Figure 12 displays the magnetic field distribution in the $YZ_{MSE}$ plane as predicted by the PINN-A1 model.

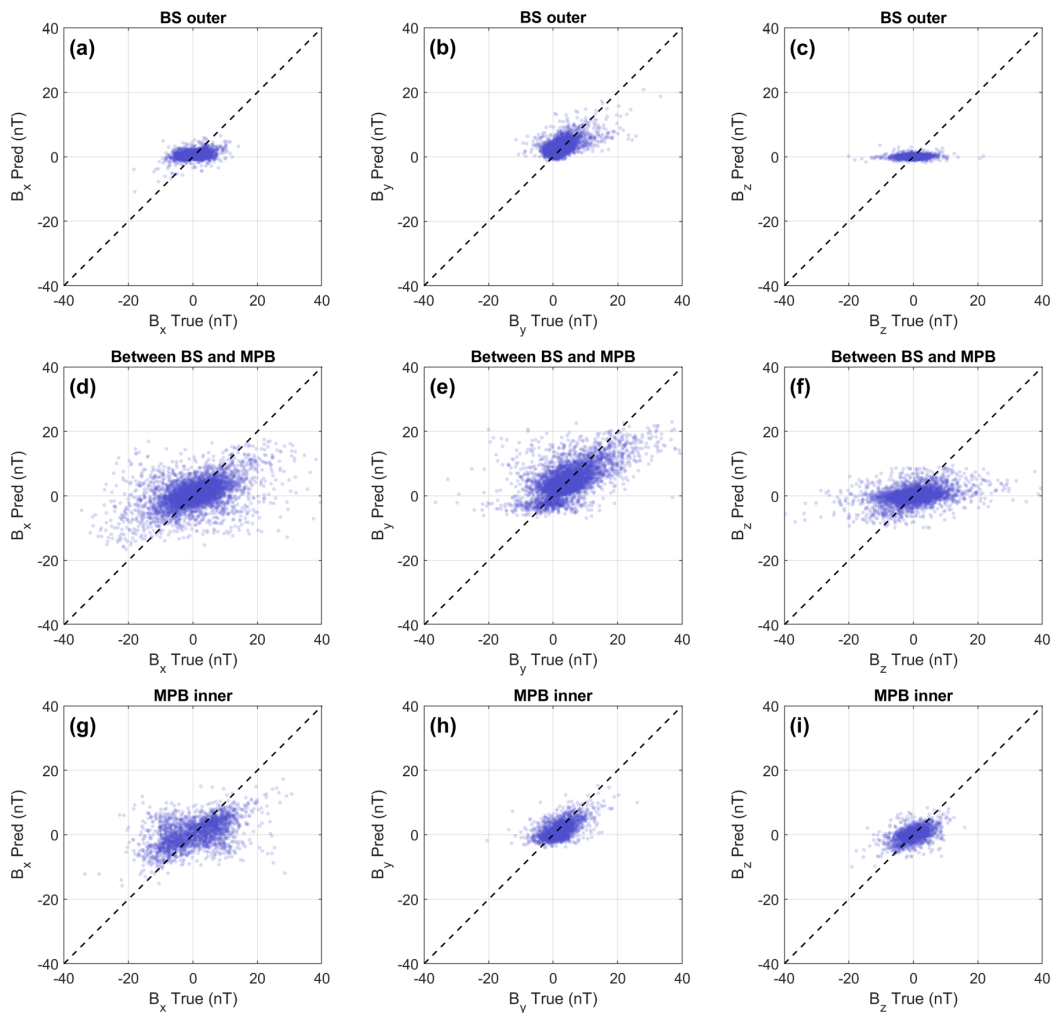

**Figure 6. Correlation between the MAVEN magnetic field observations and the PINN-A1 model predictions.** From top to bottom, the three rows show the correlation maps for the region

outside the bow shock, the region between the bow shock and the magnetic pileup boundary (MPB), and the region inside the MPB. Each panel compares the observed and PINN-A1 model predicted magnetic field components in the MSE coordinate. The black dashed line indicates the perfect agreement. From left to right, the columns correspond to the magnetic field $B_x$, $B_y$, and $B_z$ components.

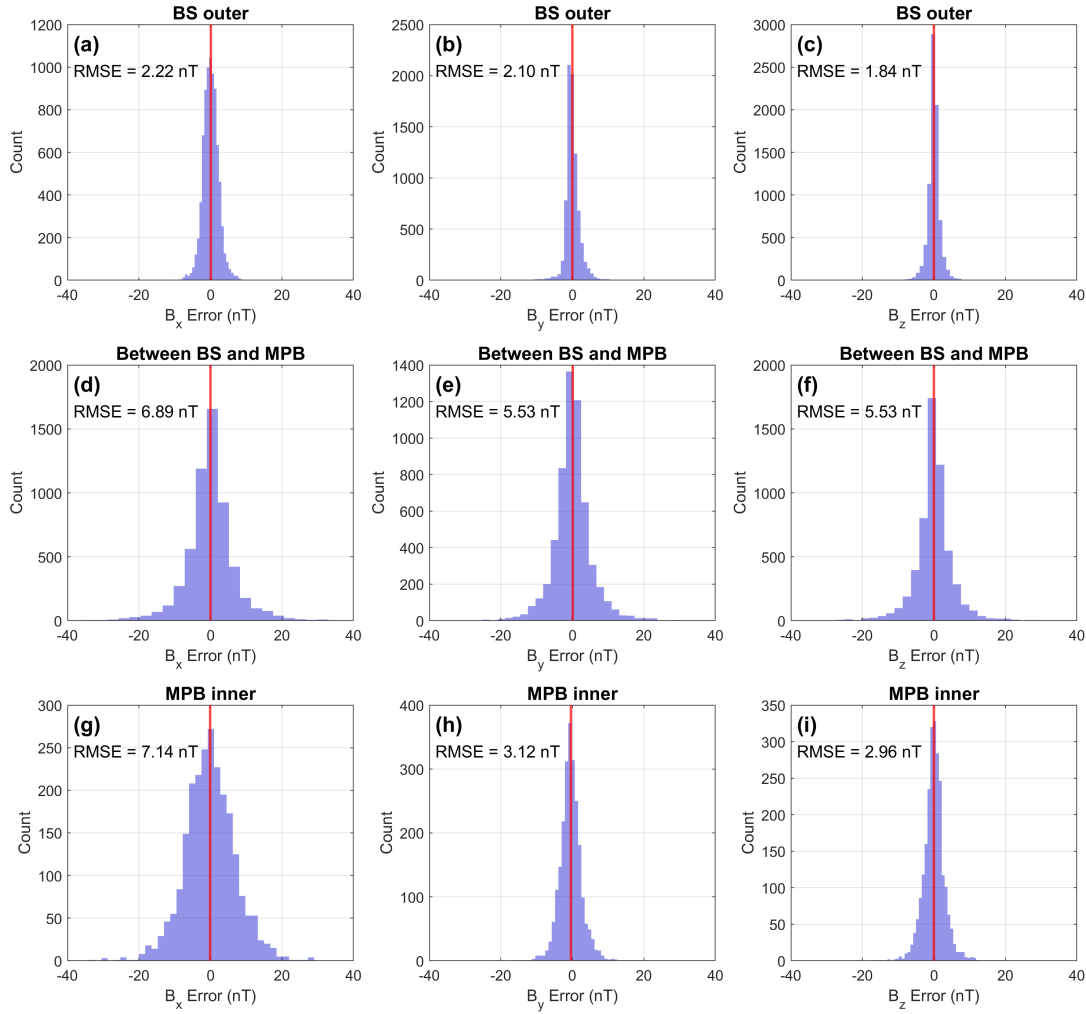

**Figure 7. Histogram of data residuals in different regions.** Similar to Figure 6 but showing the distributions of data residuals. Residuals are calculated as the difference between the observations and the PINN-A1 model predictions. The root-mean-square error (RMSE) for each magnetic field component is shown in the upper-left corner of each panel. Red vertical line indicates the zero.

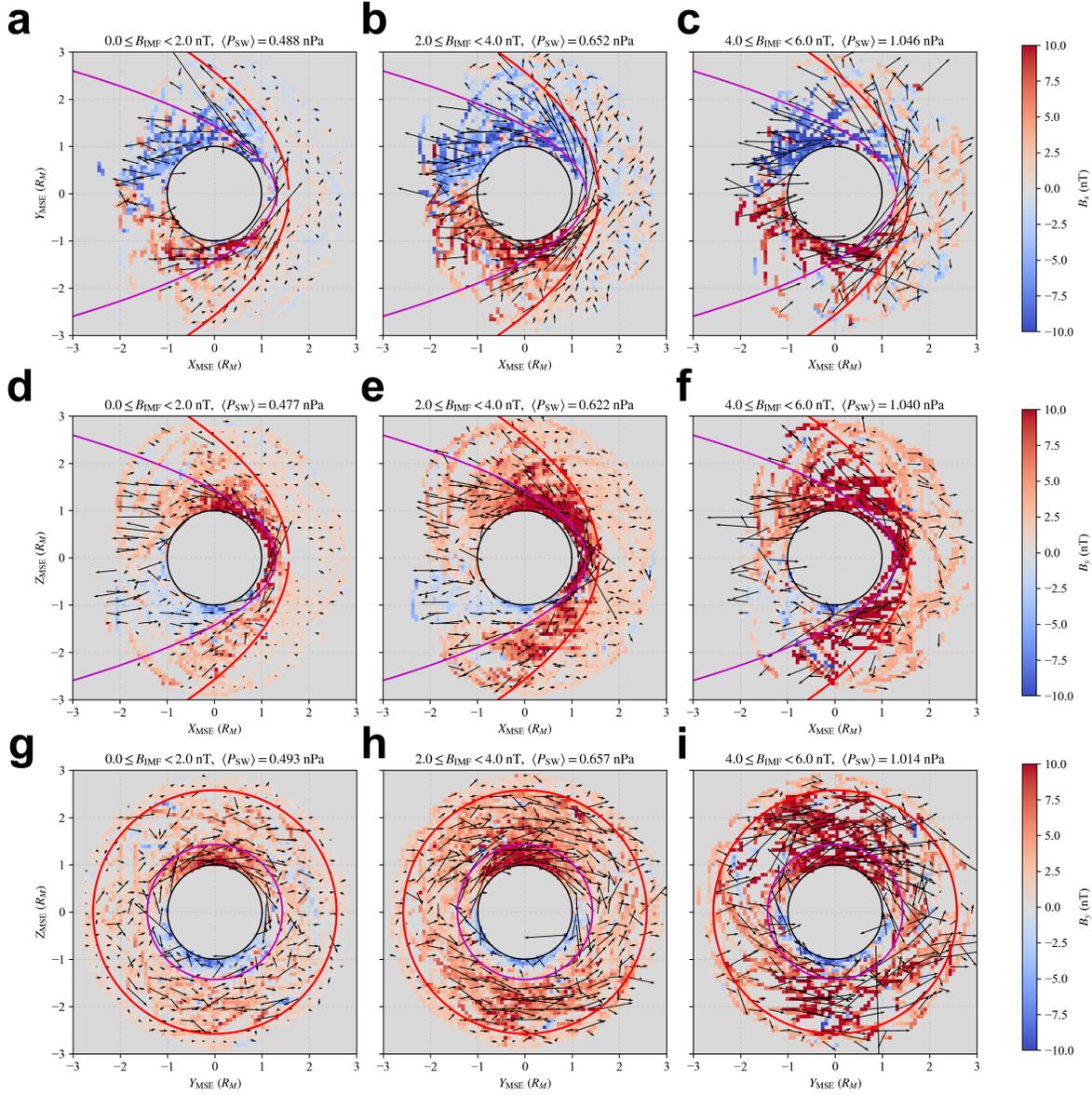

**Figure 8. Magnetic field distribution in the Martian induced magnetosphere under varying upstream IMF strengths from MAVEN observation.** Magnetic field vectors are shown in the slices of the (a-c) XY$_{MSE}$, (d-f) XZ$_{MSE}$, and (g-i) YZ$_{MSE}$ planes, respectively. Each slice is partitioned by a bin of 0.2 $R_m$. Panels from left to right correspond to upstream B$_{IMF}$ ranges of 0-2 nT, 2-4 nT, and 4-6 nT. The average $P_{SW}$ during each observation period is indicated in the subtitles. The red and magenta lines denote the shape of the bow shock and the magnetic pileup boundary (MPB) (Němec et al., 2020). Note that in the XY$_{MSE}$ plane, the color bar represents the B$_x$ component, whereas in the XZ$_{MSE}$ and YZ$_{MSE}$ planes, it represents the B$_y$ component.

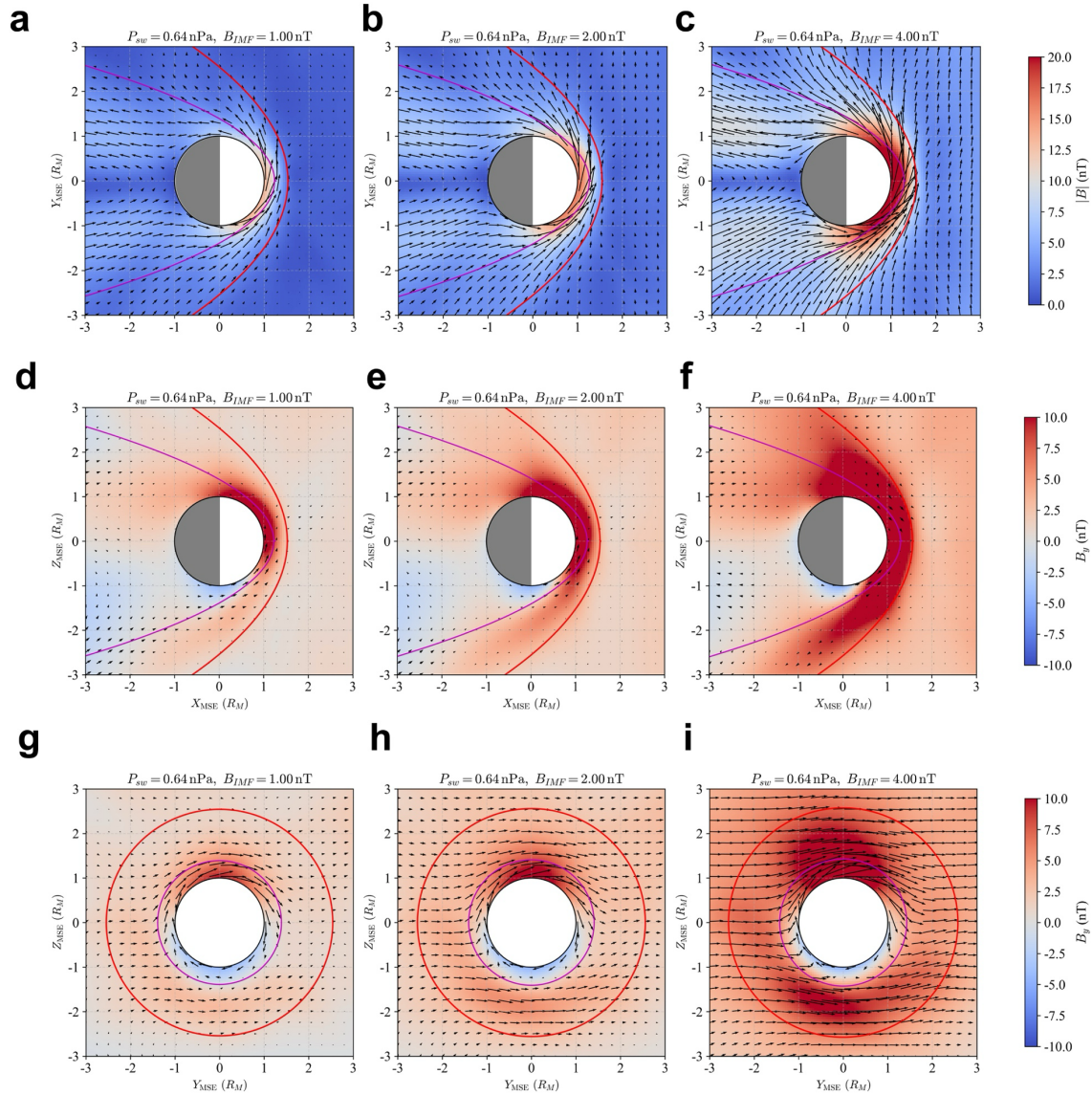

**Figure 9. Magnetic field distribution in the Martian induced magnetosphere under varying upstream IMF strengths from the PINN-A1 model without IMF replacement.** Magnetic field vectors are shown in the slices of the (a-c) $XY_{MSE}$, (d-f) $XZ_{MSE}$, and (g-i) $YZ_{MSE}$ planes, respectively. Panels from left to right correspond to upstream $B_{IMF}$ of 1 nT, 2 nT, and 4 nT. The $P_{SW}$ is fixed at 0.64 nPa in all cases. The red and magenta lines denote the shape of the bow shock and the magnetic pileup boundary (MPB) (Němec et al., 2020). Note that in the $XY_{MSE}$ plane, the color bar represents the magnetic field intensity, whereas in the $XZ_{MSE}$ and $YZ_{MSE}$ planes, it represents the $B_y$ component.

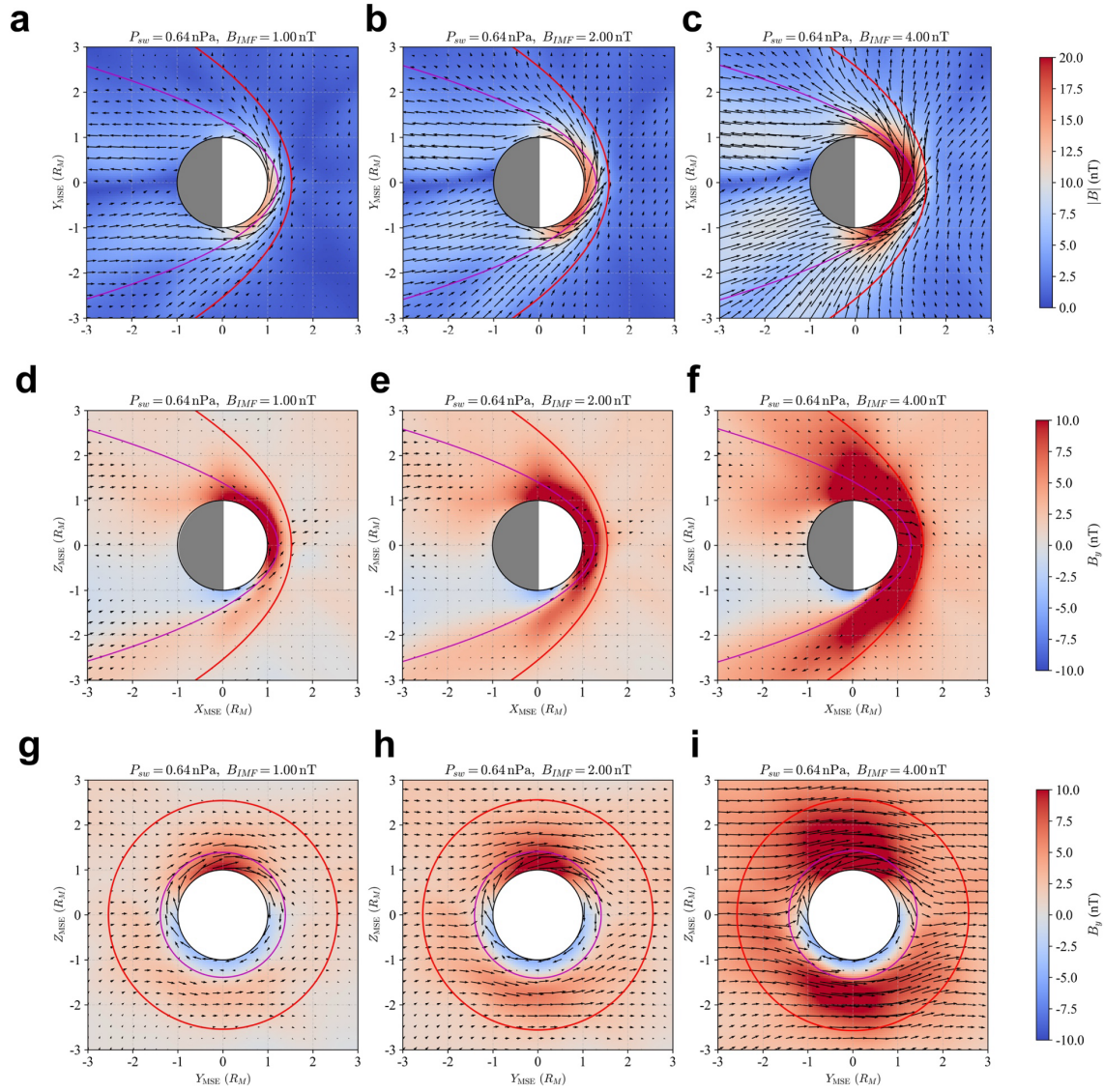

**Figure 10. Magnetic field distribution in the Martian induced magnetosphere under varying upstream IMF strengths from the PINN-A2 model without IMF replacement.** Magnetic field vectors are shown in the slices of the (a-c) $XY_{MSE}$, (d-f) $XZ_{MSE}$, and (g-i) $YZ_{MSE}$ planes, respectively. Panels from left to right correspond to upstream $B_{IMF}$ of 1 nT, 2 nT, and 4 nT. The $P_{SW}$ is fixed at 0.64 nPa in all cases. The red and magenta lines denote the shape of the bow shock and the magnetic pileup boundary (MPB) (Němec et al., 2020). Note that in the $XY_{MSE}$ plane, the color bar represents the magnetic field intensity, whereas in the $XZ_{MSE}$ and $YZ_{MSE}$ planes, it represents the $B_y$ component.

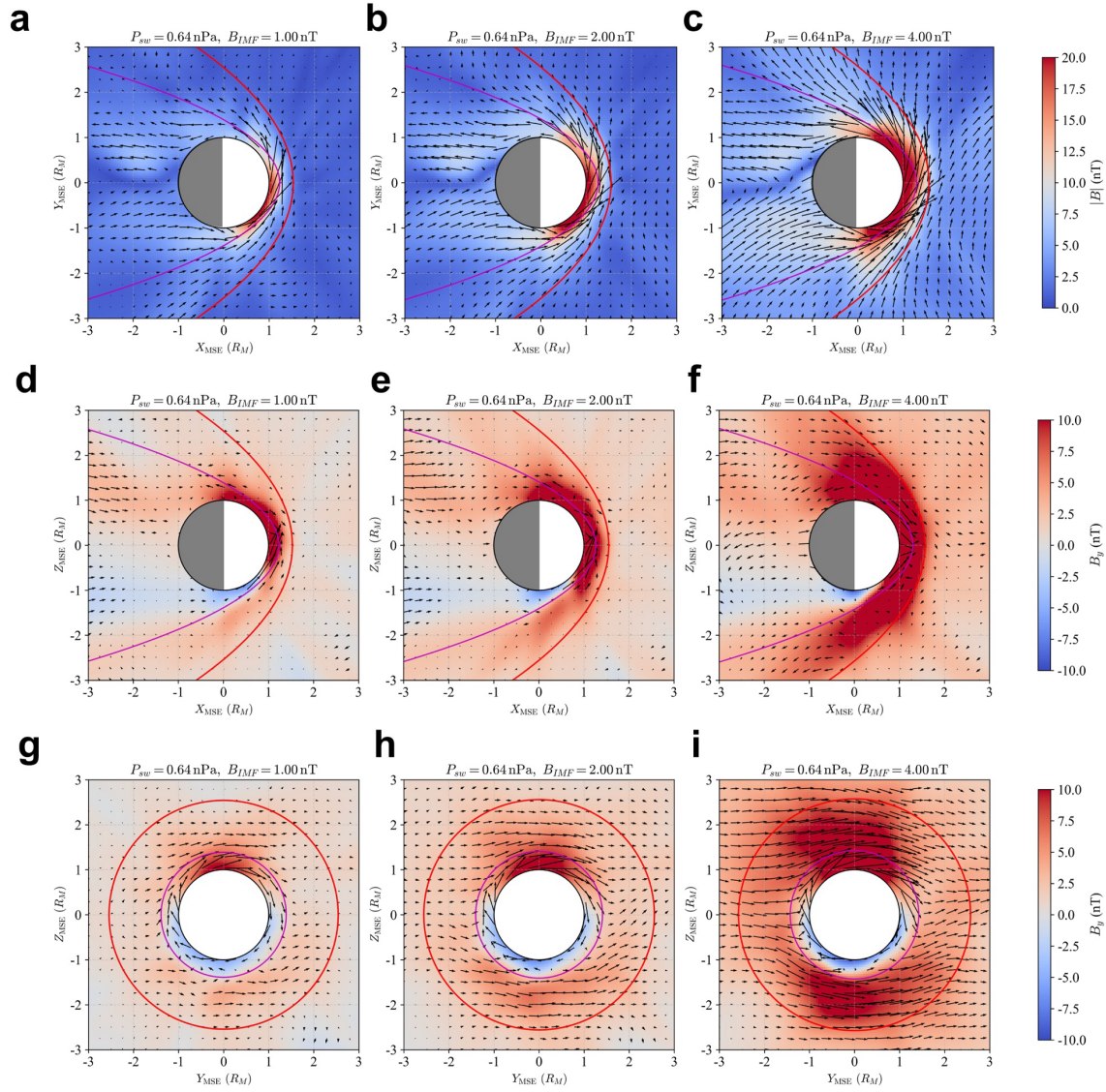

**Figure 11. Magnetic field distribution in the Martian induced magnetosphere under varying upstream IMF strengths from the PINN-A3 model without IMF replacement.** Magnetic field vectors are shown in the slices of the (a-c) $XY_{MSE}$, (d-f) $XZ_{MSE}$, and (g-i) $YZ_{MSE}$ planes, respectively. Panels from left to right correspond to upstream $B_{IMF}$ of 1 nT, 2 nT, and 4 nT. The $P_{SW}$ is fixed at 0.64 nPa in all cases. The red and magenta lines denote the shape of the bow shock and the magnetic pileup boundary (MPB) (Němec et al., 2020). Note that in the $XY_{MSE}$ plane, the color bar represents the magnetic field intensity, whereas in the $XZ_{MSE}$ and $YZ_{MSE}$ planes, it represents the $B_y$ component.

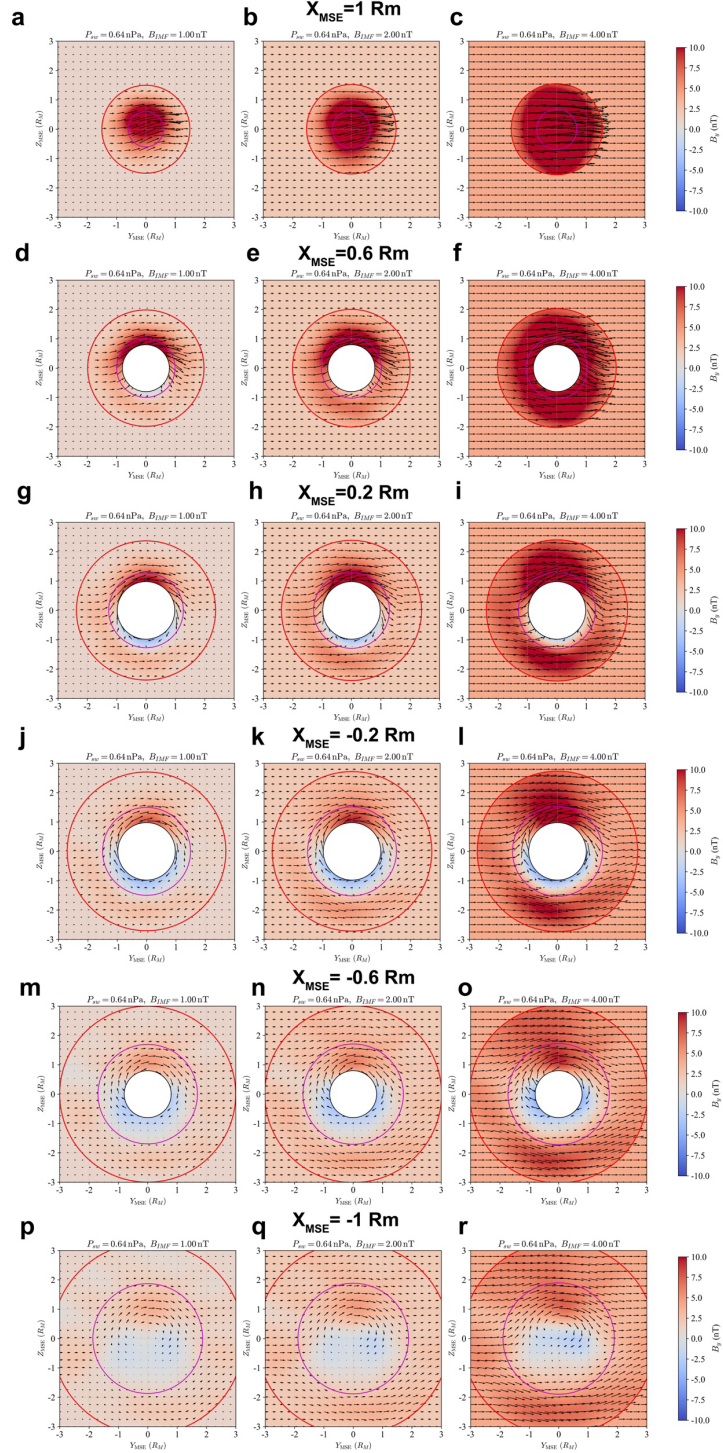

**Figure 12. Magnetic field distribution in the $YZ_{MSE}$ plane from the PINN-A1 model.** Magnetic field vectors are shown in slices of the $YZ_{MSE}$ planes. Panels from left to right correspond to upstream $B_{IMF}$ values of 1 nT, 2 nT, and 4 nT. The $P_{SW}$ is fixed at 0.64 nPa for all cases. Panels from top to bottom step in the -X direction from 1 $R_m$ to -1 $R_m$. The red and magenta lines denote the shape of the bow shock and the magnetic pileup boundary (MPB). The color bar indicates the intensity of the $B_y$ magnetic field component.

**Acknowledgments**

This work was partially supported by the MAVEN Project, NASA grants 80NSSC23K1125, 80NSSC23K0911, 80NSSC24K1843, and the Alfred P. Sloan Research Fellowship. We thank Yaxue Dong and Xiaohua Fang for their helpful discussion.

MAVEN magnetic field data are publicly available through https://lasp.colorado.edu/maven/sdc/public/data/sci/mag/l2/ (Connerney, 2023). MAVEN SWIA data are publicly available through https://lasp.colorado.edu/maven/sdc/public/data/sci/swi/l2/ (Halekas, 2017). The PINN models developed in this study have been deposited in a Github repository and are openly available at https://github.com/gaojiawei321/Mars_PINN_Mag.